\documentclass{rsifpublic}

\usepackage{graphicx}
\usepackage{amsmath}
\usepackage{amsfonts}  
\usepackage{amssymb}
\usepackage{epstopdf} 
\usepackage{moreverb}
\usepackage{psfrag}

\newcommand{\real}{\rm real}

\begin{document}

%
%
%

\runningheads{R.Hanel}{Attractors \& Cell Differentiation}

\begin{topmatter}{

\title{Systemic stability, cell differentiation, and evolution}
{A dynamical systems perspective}

\author{Rudolf Hanel\corrauth}
\address{Section for Science of Complex  Systems, CeMSIIS, Medical University of Vienna, Spitalgasse 23, A-1090, Vienna, Austria}

\begin{abstract}
Species or population that proliferate faster than others become dominant in numbers. 
Catalysis allows catalytic sets within a molecular reaction network to dominate the non catalytic 
parts of the network by processing most of the available substrate. 
As a consequence one may consider a '{\em catalytic fitness}' of sets of molecular species.
The fittest sets emerge as the {\em expressed} chemical backbone or sub-network of 
larger chemical reaction networks employed by organisms. 
However, catalytic fitness depends on the systemic context and the stability of systemic 
dynamics. Unstable reaction networks would easily be reshaped or destroyed by
fluctuations of the chemical environment. 
In this paper we therefore focus on recognizing systemic stability as an evolutionary selection criterion.
In fact, instabilities of regulatory systems dynamics become predictive for associated evolutionary forces 
driving the emergence large reaction networks that avoid or control inherent instabilities. 
Systemic instabilities can be identified and analyzed using relatively simple mathematical random networks models 
of complex regulatory systems. Using a statistical ensemble approach one can identify 
fundamental causes of instable dynamics, infer evolutionary preferred network properties, 
and predict evolutionary emergent control mechanisms and their entanglement with cell differentiation processes. 
Surprisingly, what systemic stability tells us here is that cells (or other non-linear regulatory systems) 
never had to learn how to differentiate, but rather how to avoid and control differentiation.
For example, in this framework we can predict that regulatory systems will evolutionary favor networks where the 
number of catalytic enhancers is not larger than the number of suppressors.
\end{abstract}

}
\end{topmatter}

\corraddr{(rudolf.hanel@meduniwien.ac.at)}



\vspace{0.3cm}

\section{Introduction}

Cellular life as we know it today has evolved from chemical processes. Several crucial {\em inventions} 
had to be made on this way, such as {\em replication} and {\em self-replication}, hereditary information, 
self-maintenance, and differentiation. In this sense contemporary life-forms represent their evolutionary 
history in ``stratigraphic'' layers of chemical processes that obviously worked together well enough to be 
transmitted down the ages, molding hereditary information into the {\em molecular regulatory networks} (MRN) 
of cells. This indicates {\em systemic stability}, a ``constancy of form'', as a crucial evolutionary 
selection criterion on a systemic level. Which formal reasons support this conjecture?

It is an open question to which extent chemical and biological evolution would follow a similar 
evolutionary path under similar conditions. Clearly, there is a strong random element in evolutionary 
processes and rewinding the evolutionary clock,   
rerunning earth history, would lead to an ``alternative'' biology, different from the particular one we know. 
Nevertheless, the direction of evolution is not completely arbitrary but seems to follow certain principles of 
organization, which emerge from the complex chemistry and physics of biological processes (\cite{LenskiEColi,AlberchMonsters,SCMorris,FontanaPlayedTwice}).  
Identifying and understanding such principles forms the fundamental challenge of a {\em general theory} of {\em systems biology}. 

Some principles governing complex systems dynamics need to be complementary to Darwinian evolution, 
which can only act once hereditary information exists. Copy processes and self-maintaining organization \cite{FontanaPlayedTwice} 
necessarily belong to the repertoire of pre-Darwinian evolution. For instance, there is evidence that Ribosomes began as 
self replicating molecular machines \cite{ribosomes}, which gave other molecules opportunity to hijack Ribosomes for their own 
replication, which in turn may explain the fact that all organisms we know today replicate using Ribosomes.
Catalysis on the other hand, acting as the {\em proliferative fitness} of sub-MRN, 
is key to understanding the emergence of ``self'' (the organism) in self-maintaining organization   
that propagates its functional form {\em and} dynamically adapts by running differentiation programs.  

Almost by definition, we believe that we have understood something about the world, if we can reproduce it in some sense. 
Either by engineering, or by constructing a mathematical model within an abstract framework, an algorithm, that by using 
{\em correspondence principles} interprets and predicts {\em relevant parts} of the phenomenology of interest.  
In this spirit we are going to show how {\em systemic stability} can be investigated using mathematical models 
and serves as a principle of (pre-Darwinian) systemic organization that 
continues to govern the evolution of contemporary organisms. In fact this principle can be applied to understand
the mechanism underlying one of the hot topics of {\em regenerative medicine}: 
{\em induced pluripotent stem cells} (iPS); triggering re-differentiation of cells into different tissues such as cartilage, 
muscles, lungs, or neurons. From a formal perspective, the possibility of inducing re-differentiation by perturbing a cell with 
specific molecular vectors of so called reprogramming factors (e.g. \cite{Yamanaka2006}) is an immediate consequence 
of the correspondence of cellular regulatory processes with complex dynamical systems (compare Fig. 1).  

We use the {\em correspondence} of non-linear systems and MRN to identify cellular differentiation status 
with dynamical system attractors. This allows us to characterize the abstract principle of organization 
shaping MRN dynamics at a systemic level. Our analysis shows that in a fluctuating environment the natural thing for cells 
(or any other self-perpetuating complex non-linear adaptive system) to do, is to switch between different states of differentiation. 
In other words, cell differentiation comes for free! As a consequence, what cells had to learn was {\em not} to {\em spontaneously
differentiate} under the whims of a fluctuating environment. To achieve this, as already earlier work of ours indicated \cite{Poech10,Poech12}, 
cells had to learn to maintain a tight control of the rates at which functional molecules become degraded back into substrate. 
This is especially true for functional molecules, such as proteins, that degrade very slowly (or not at all) under typical physical conditions eco-systems we know of provide. 

As we will show below, the need for controlling degradation rates has little to do with adjusting the abundance of molecules. 
Intuitively, one would suspect that higher degradation rates imply lower abundance of molecules (a lower chance of being expressed). 
On a systemic level this turns out to be wrong in general, demonstrating the context sensitive role molecules play within regulatory systems.

The reasons why cells require {\em degradation rate control} (DRC) systems are rooted deeply in mathematical properties of complex 
dynamical systems that link the likelihood of differentiation events to the molecular diversity of a system. If random 
differentiation comes for free, then the challenge for multicellular life is to control the natural inclination of cells to differentiate 
spontaneously, by evolving cell differentiation control (CDC) systems. The easiest way for CDCs to emerge is simply by involving early 
successful (pre-Darwinian) DRCs into the (later Darwinian) evolution of MRN.

Biological evidence corroborates these conjectures. For instance, one DRC all eukariota employ, is the 
Ubiquitin Proteasome System (UPS) for degrading proteins \cite{Ciech1998,Gsponer2008}. 
Ubiquitin and ubiquitin-like proteins have a broad spectrum of tasks. One task is to tag proteins for 
degradation by proteasome. Proteasomes (but not Ubiquitin) can also be found in all archaea and some prokariota, 
for instance in Mycobacterium tuberculosis \cite{MycBT}, where Ubiquitin is substituted by a 
different species of molecules (Pup). Proteasomes are important evolutionary old and highly conserved components of DRC, 
though not the only ones (e.g. \cite{altdegsys}). UPS also plays a key role in cell 
differentiation \cite{Reavie2010,Ishino2014} and stem cell pluripotency \cite{Buckley2012}. 
This demonstrates the adequacy of the hypothesis that systemic stability plays a fundamental role governing evolution
and that DRC systems in general are likely to be linked to cell differentiation programs.

How can we utilize systemic stability as a {\em principle of evolutionary organization}?
For this we use a modeling approach we have already introduced in earlier work \cite{Stokic2008,Poech10,Poech12}.
This approach is based on a tradition of mathematical systems biology to use differential equations 
for modeling complex biological systems (e.g. \cite{Hofbauer1998}), and to exploit 
correspondence principles between the formal framework and the observable phenomenology. 
We combine this approach with a statistical ensembles approach. Ensembles can be used (just like in statistical physics) 
to obtain results - not for any system in particular, but for what is called a {\em typical} system (the majority of similarly behaving systems). 

We briefly sketch the mathematical approach and summarize some of our previous work. 
We discuss cell differentiation as a fundamental emergent property of MRN
and the predictive role systemic stability may take in a general theory of systems biology, 
that may extend to evolutionary non-linear complex regulatory systems and their differentiation dynamics in general. 
In particular, we present new results on systemic stability properties of MRN. We observe a surprising failure of degradation rates to 
be predictive for the expression level of a molecule type and infer (from stability considerations) that evolution is likely to prefer MRN where positive catalytic links do not outnumber negative links which again forms a testable hypothesis.

\begin{figure}[t]
\begin{center}
		\includegraphics[width=0.8\columnwidth]{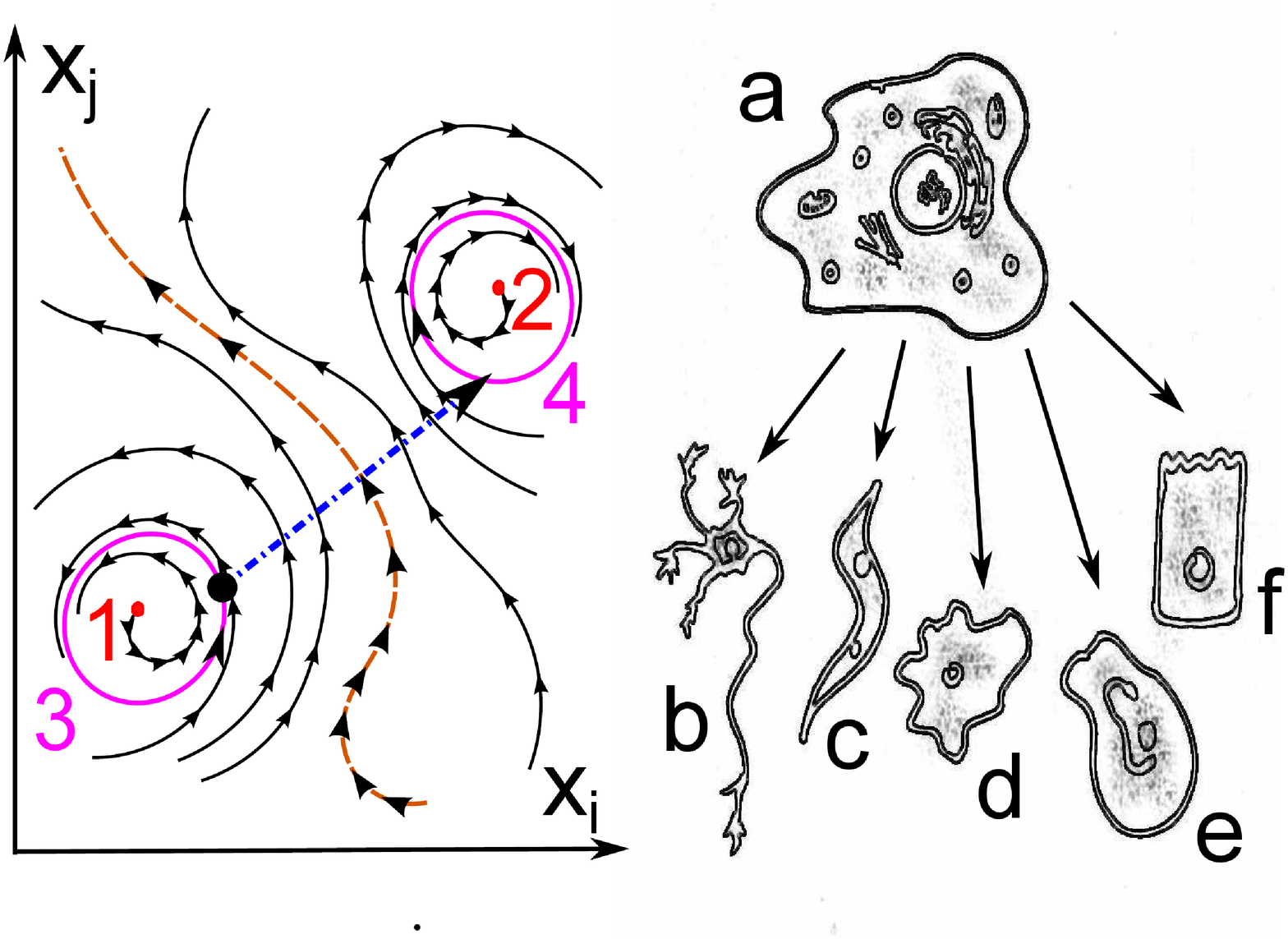}
\end{center}
	\label{fig:cellsandattractors}
	\caption{{\bf Cell differentiation and attractors}: We show, as a cartoon, (right) a stem cell (a) differentiating into different cell types; e.g. (b) neurons, (c) muscle cells, (d) macrophages, (e) red blood cell, (d) epithelial cell. Cells can differentiate into different cell types, using the dynamics of one and the same molecular regulatory system. Non-linear dynamical systems of many variables $x_i$ can possess multiple distinct attractors. In the cartoon (left) there are two instable fixed points (1,2) and two stable limit cycles (3,4). The dashed line separates the basins of attraction of the stable limit-cycles 3 and 4. 
The dash-dotted arrow corresponds to a reprogramming vector that allows to push the system from limit cycle 3 to limit cycle 4.   
Distinctly differentiated cells are the physical representation of distinct attractors of the organisms MRN dynamics.
}	 	
\end{figure}

\subsection{Differentiation status and attractors}

For deriving conclusion about systemic organization one requires a systemic framework to begin with and 
correspondence principles that allow us to identify features of the formal framework with observable phenomenology! -
Cells are not like houses. Their parts (e.g. molecules such as proteins) not only get substituted 
after being damaged (although molecular quality control is an important feature of contemporary organisms). 
In particular, regulatory molecules typically get produced from substrate, get used, and are again degraded 
back into substrate, once the job is done. Cells are complex dynamical systems formed by a structurally complex and 
chemically diverse molecular machinery. 
The interior of cells is separated from the environment by a cell membrane,
may be compartmentalized (e.g. nucleus \& cytosol in eukariota), and contains the 
genome that encodes heritable information in form of genes. Genes can be activated, mRNA gets transcribed  
from genes, mRNA gets translated into proteins, and proteins may play a structural, functional or regulatory role.
Some proteins, called transcription factors, can regulate the transcription rate of genes by either promoting or suppressing binding 
of the molecular transcription machinery to the particular gene.
The system of chemical processes that regulate the abundance of molecular agents 
we refer to as the MRN of the cell. 

If we think of cells as the crowded molecular environments that they turned out to be \cite{crowding}, 
then, at a very fine resolution, cells formally look like Chimeras, partly boolean networks, partly reaction-diffusion systems, and partly cellular-automaton. Merging these mathematical concepts into hybrid models may become unavoidable in order to understand systemic details. However, this is not our aim here. We attempt to understand the general implications of systemic stability on emergent control mechanisms, MRN topology and cell-differentiation. 

We can simplify the situation by taking a step back to look at cells at a coarser spatial and temporal resolution.
At this resolution location becomes a discrete property, e.g. $i=(I\kappa B,\ {\rm nucleus})$ and 
$i'=(I\kappa B,\ {\rm cytosol})$. This allows us to think of the variable $x_i(t)$ simply as the number of molecules of 
type $i$ at time $t$. We collect all considered types $i$ in a collection $I=\{1,2, 3,\cdots,N\}$, called the index set.
Counting measures are always non-negative, 
i.e. $x_i(t)\geq 0$ for all $i$ and $t$ which is of crucial importance for MRN dynamics. 
$x=(x_1,x_2,\cdots,x_N)$ is a list of {\em molecular abundances} and {\em gene-activities}. 
The {\em diversity} $N$ is the number of distinguishable molecular types and genes. 
We also go from asking whether something happens (e.g. a gene gets expressed) or not, to asking at which rate this happens. 
The dynamics of $x_i(t)$ then can be described by ordinary differential equations (ODE) of the form
\begin{equation}
	\dot x_i(t)=F_i(x(t))\ +\ {\rm noise}\,,
	\label{NLDGL}
\end{equation}
where $\dot x_i(t)=d x_i(t)/dt$ is the first derivative of $x_i(t)$, which measures how much $x_i(t)$ changes per time unit at time $t$.
$F_i$ is a function that depends on the abundance of all involved molecule types and in general is non-linear.
The non-negativity of $x_i$ additionally imposes a non-linear boundary condition we refer to as positivity-condition (PC)
\begin{equation}
x_i(t)\geq 0\, .
\label{poscond} 
\end{equation}
For any two molecule types $(i,j)$ we can draw the dynamics $(x_i,x_j)$ as a line on a sheet of paper, a two dimensional coordinate system.
For ``drawing'' $x$ we need $N$ dimensions.

It is well known that the dynamics of high-dimensional non-linear processes (disregarding noise) concentrates in 
particular regions on this $N$ dimensional ``sheet of paper'', where $x(t)$ remains for ever (if not exogenous 
influences perturb the system). Such regions are called {\em attractors} of the dynamics, and can be thought of as 
the valley floors in a wrinkled hilly landscape. The simplest attractor is a single point, a so called {\em fixed point}. 
A periodic attractor is called a {\em limit cycle}. Depending on the initial condition $x(t_0)$, the initial chemical 
composition of a cell at time $t_0$, $x(t)$ may end up ($t_0<t\to\infty$) in different distinct attractors. Each 
attractor possesses a {\em basin of attraction}. Such a basin consists of all possible initial conditions $x(t_0)$ 
that lead into the same attractor. High dimensional non-linear systems may possess large numbers of attractors. 

Noise (compare Eq. (\ref{NLDGL})) smears out attractors and occasionally
a fluctuation may push $x(t)$ from one basin of attraction into an adjacent basin and the dynamics will start to follow the new attractor.
If the dynamics of molecular abundance in cells is (at least approximately) described by equations of the form Eq. (\ref{NLDGL}),
then different attractors correspond to different ways a cell can express its MRN. 
As a consequence, one can identify distinct attractors of the dynamics with distinctly differentiated cells, 
establishing a {\em correspondence principle} with profound implications. 
For instance, re-differentiation of cells can occasionally be triggered by stochastic molecular fluctuations, 
which may result from perturbations in the chemical environment, other forms of stress, or signals from other cells. 
Alternatively, one might {\em designed} such perturbations to push the MRN dynamics from one basin of attraction to an adjacent basin. 
Reprogramming vectors that re-differentiate cells (e.g. the iPS cell generating Yamanaka vector \cite{Yamanaka2006}) exemplify 
such designer perturbations.

In this precise sense differentiated cells {\em are} physical representations of MRN attractors. Only since very recently one begins to discover, that by applying reprogramming vectors, cell types can be induced that so far have never been described for the respective organism (e.g. \cite{pluripot,pluripot2}). This substantiates the {\em correspondence principle} between attractors and the differentiation status of cells. It also indicates that multicellular organisms usually avoid to use their full repertoire of differentiated states. - In order to explain why precisely cells require DRC systems we unfortunately have to dig a little deeper into the mathematical properties of dynamical systems.

\subsection{The mathematical approach}

Important components of cellular MRN are transcription factors, 
mostly proteins that have the function of catalysts (enhancers/suppressors) that promote/suppress the transcription of other genes. 
Chemical reaction rates are either boosted in the presence of catalytic enhancers, or get very low in the presence of suppressors. 
The enhanced processes form regulatory sub networks that process most of the available substrate. 
This allows particular sub networks to dominate 
other possible processes (which add to the background noise). 
In this idealization $F_i(x)$, Eq. (\ref{NLDGL}), becomes linear:
\begin{equation}
	F_i(x)=J_i+\sum_{j\in I} A_{ij}x_j-D_i x_i\,.
	\label{LINEQ}
\end{equation}
$A_{ij}$, the proportionality constant between $F_i$ and $x_j$, is called the (weighted) adjacency matrix of the {\em catalytic} 
molecular regulatory network (cMRN). If $A_{ij}\neq 0$ we speak of a link from $j\to i$. 
If $A_{ij}=0$, there is no link from $j$ to $i$. 
If $A_{ij}>0$, then $j$ promotes the production of $i$, if $A_{ij}<0$, $j$ suppresses it. 
$J_i$ model input/output-flux and/or base-line production rates of molecules of type $i$. 
The {\em degradation rates} $D_i$ quantify the fraction of molecules $i$ that are degraded within a time unit. 
Alternatively, we can write Eq. (\ref{LINEQ}) in the following way: $F_i(x)=\sum_{j\in I} B_{ij}(x_j-x^*_j)$. 
In this notation catalytic interactions $A$ and degradation rates $D$ are considered together in a single matrix $B$, with
$B_{ij}=A_{ij}$ for $i\neq j$ and otherwise $B_{ii}=A_{ii}-D_i$. Formally, $x_i^*$ is a fixed point of the 
dynamics, with $J_i=-\sum_{j\in I} B_{ij}x^*_j$. However, due to PC, $x^*$ (stable or unstable) is only {\em accessible} to the dynamics, 
i.e. $x=x^*$ is possible, if $x^*_i\geq 0$ for all $i=1,\cdots,N$, which is not a priorly guaranteed. 

While Eq. (\ref{LINEQ}) is formally linear, the PC, Eq. (\ref{poscond}),
introduces a non-linear constraint. 
PC becomes active whenever some molecule $i$ attains zero abundance and would continue to become negative if this were possible. 
At any time $t$ this divides the considered molecules $I=\{1,\cdots,N\}$ into two disjoint index collections $I^+$ and $I^-$. 
The {\em active} molecules $I^+(t)=\{i\in I|x_i(t)>0\}$; and the {\em inactive} molecules $I^-(t)=\{j\in I | x_j(t)=0\}$, subject to PC.
We call the catalytic sub-network of active molecules ($i\in I^+$) the {\em active},  
or synonymously, the {\em expressed} regulatory network.

Typically non-linear dynamics are {\em multi-stable}, i.e. several distinct $I^+$ exist for a 
fixed MRN so that the regulatory dynamics of the distinct expressed networks approaches a stable fixed point. 
For limit cycles the sets $I^{\pm}(t)$ change at discrete switching times,
when some type of molecules becomes inactive (zero abundance) or active (reproduced). 

Between switching events $I^{\pm}$ are constant.
Moreover, for any particular $I^+$ the active molecules $i\in I^+$ follow the linear dynamics 
$\dot x_i=J_i+\sum_{j\in I^+} A_{ij}x_j-D_i x_i$. 
The {\em active} dynamics of $x(t)$ is governed merely by matrix elements $A_{ij}$ 
restricted to $i,j\in I^+$ (the active sub-matrix of $A$). 
Whenever a switching event occurs (in a limit cycle or induced by perturbations) the dynamics $x(t)$ {\em sequentially} switches from one 
active subsystem to another one, in the attempt of finding a new stable, accessible fixed point. 
{\em Switching} is also the mechanism that allows limit cycles to emerge and MRN to operate close to the
so called {\em edge of chaos} (maximal Lyapunov exponent $\lambda\sim0$)
by balancing the time the limit cycle spends passing through stable and instable sub-system dynamics \cite{Poech12}.  

Sequential linear (SL) systems are related to so called {\em piecewise linear systems}, 
e.g. Glass-Kauffman networks \cite{GlassKauffman1973}. Unlike piecewise linearity, sequential linearity is a consequence of PC, 
Eq. (\ref{poscond}), and not a consequence of rate-parameters changing abruptly at critical concentration levels.

Systems governed by linear equations can be analyzed using powerful methods from {\em linear algebra}. 
For instance, for matrices $A$ there exist distinct constants $\eta$ called {\em eigenvalues} (which in general are complex numbers) 
and associated {\em eigenvectors} $v$ so that $\eta v=Av$, i.e. $\eta v_i=\sum_j A_{ij} v_j$. 
For eigenvectors matrix $A$ behaves like the scalar $\eta$.
Therefore, in Eq. (\ref{LINEQ}), the maximal real part of $\eta$, 
will determine how the abundances $x_i$ behave, and can be identified with $\lambda$, the maximal Lyapunov exponent of the subsystem, measuring the stability of the (sub-)systems dynamics. 
If $\lambda>0$, then small perturbations of the dynamics will grow exponentially 
with time and fixed points will be unstable; if $\lambda<0$ perturbations will fade exponentially and fixed points are stable.

\subsection{Ignorance, statistical ensembles, and typical systems}

While SL systems can be used successfully for modeling small modules (relatively independent sub networks) 
explicitly (compare e.g. \cite{Poech12} Fig.(2)), for larger systems, typically ignorance of the exact topology 
of the cMRN and of the system parameters leads 
to problems similar to the ones encountered with applying Lotka-Volterra equations to modeling real-world predator-prey relationships 
(compare \cite{Mason2008}). The question arises how mathematical models may still 
inform us on properties of the dynamics of large MRN. 

Statistical physics taught us how to deal with ignorance by using a statistical ensemble approach to 
model the behavior of {\em typical} systems rather than the exact behavior of any particular system. 
Suppose, what we know about a system are the properties ${\cal P}$. Any system that possesses ${\cal P}$ is 
{\em admissible}, since we can not discriminate between systems sharing ${\cal P}$. To get a fair picture of the
expected systemic behavior one may simply work with the collection of all admissible systems sharing ${\cal P}$, 
called a statistical ensemble. 
${\cal P}$ may include the knowledge that, on average, each molecule type $i$ gets influenced by $\bar k$ 
other molecule types $j$, or that a fraction $p$ of those links are catalytic enhancers, and a fraction $(1-p)$ 
are suppressors. We might also know the distribution of link weights $A_{ij}$, degradation rates $D_i$, and 
estimates of global fixed point values $x^*_j$. 
For simplicity we consider $D_i$ to be evenly distributed in the interval $\bar D(1-\delta_D) <D_i<\bar D(1+\delta_D)$.
Once ${\cal P}$ has been fixed, one can start to sample SL models from the corresponding ensemble,
simulate the dynamics of each sample, and measure properties of its dynamics. 
In this way we obtain a statistics of 
dynamical characteristics of systems constrained by ${\cal P}$ and
can explore how changes of ${\cal P}$ affect the {\em typical} dynamics, providing a quantitative, 
rational basis for inductive reasoning about complex systems dynamics.

\section{Results}

We have analyzed such ensembles in earlier work \cite{Stokic2008,Poech10,Poech12},
both for Erd{\"os}-R{\'e}nyi and scale free network topologies. We used $\delta_D=0$, 
the $\bar k N$ non-zero $A_{ij}$ were drawn from a Gaussian distribution with zero mean ($p=0.5$) and 
unit variance ($\sigma_A=1$). Under those conditions we found that within a well defined range 
$D^-<\bar D<D^+$ the maximal Lyapunov exponent $\lambda$, more precisely its ensemble average $\langle\lambda\rangle$, 
approaches zero. The dynamics operates at (or close to) the {\em edge of chaos}, in this $\bar D$ range. 
Limit cycles operate at the {\em edge of chaos} ($\lambda\sim 0$). 
%
\begin{figure}[t]
\begin{center}
		\includegraphics[width=8cm]{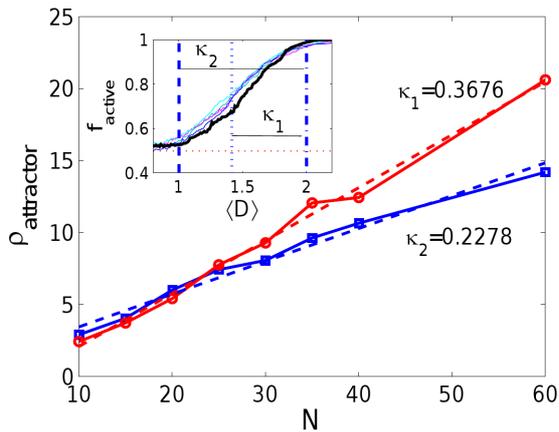}
\end{center}
	\label{fig:rhoAttractor}
	\caption{{\bf Critical degradation rate density}: For fixed average connectivity $\bar k=\langle k\rangle$
	the main critical degradation rates $D_{1/2}<D^-<D^+$ (see inset vertical lines dashed ($D_{1/2}$), 
	dotted ($D^-$), and dash-dotted ($D^+$)) are constants. The number of times 
	the active network, and therefore the attractor of the dynamics, 
	typically changes in the bounded average decay rate range $D_{1/2}<\bar D<D^+$. 
	In this range (see inset; $\bar k=4$ and $N=25,30,35,40$ and $N=60$ (black)) the fraction of the active network 
	$f_{\rm active}=N_{\rm expressed}/N$ decreases from $1$ to $1/2$ 	and the system has to repeatedly change attractors as $\bar D$ decreases. 
	The density $\rho_{\rm attractor}$ of critical degradation rates $D_{\rm crit}$ in the range has been 
	estimated for the ranges $D^-<\bar D<D^+$ (red circles) and $D_{1/2}<\bar D<D^+$ (blue squares). 
	For the estimates we count the number of times the active set-size changes by at least one node for a 
	degradation rate increment $\Delta D=0.01$ averaged over 	$50$ samples of systems with $\bar k=4$ and $N=10,15,...,40,60$.   
}	 	
\end{figure}
%
The upper and lower critical degradation rates can be estimated by 
$D^+\sim\sqrt{\bar k}$ and $D^-\sim\sqrt{\bar k/2}$, where $\bar k$ is the average degree of the network. 
For $\bar D>D^+$ typically the fixed-point $x^*$ (if accessible) becomes stable ($\lambda < 0$).
For $\bar D<D^-$ systems start to become fully unstable ($\lambda > 0$). 
$D^-$ marks a sharp transition of systems into instability at a typical expressed network size 
of $N/2$.. 

We understand now that this sharp transition at $D^-$ is a consequence of 
a sufficiently large diversity $N$ {\em and} using random initial conditions, $x(t_0)$, for each 
sampled system. However, when degradation rates change dynamically, a cell does not re-start 
its dynamics with a new random initial condition $x(t_0)$ every time this happens. 
To understand the effect of changing degradation rates on the stability of a MRN, we have to follow the 
dynamics of a system sampled from the ensembles as we vary $\bar D$. 
We start with $\bar D>D^+$ and choose initial conditions close to the fixed point 
$x^*$, which we assume to be stable and accessible. We simulate the system over a time $T$ that allows the 
dynamics to approach an attractor. We use the endpoint $x(T)$ as the initial condition for the next 
simulation of the system, where we decrease $\bar D$ by a small increment $\Delta\bar D$, and continue 
simulating the system for another time-span $T$, etc.. This sampling method has another advantage. 
We escape the necessity of extensively searching for {\em all} attractors by varying $x(t_0)$, which is a computationally 
expensive task (the computing time grows exponentially with $N$), corresponding to finding 
{\em all} (also non standard) differentiation states that can be induced (compare \cite{pluripot,pluripot2}).

When studying the stability of systems ($N\sim 50$) this way, $D^-$ marks the beginning of a 
transition into instability where stable systems remain to exist down to $D_{1/2}\sim D^+/2$. 
Around $D_{1/2}$ the size of the expressed network size $N_{\rm expressed}=|I^+|$ levels out at $N/2$.
For $N\sim 50$ we see that for $\bar D\sim D^-$ we get $N_{\rm expressed}\sim 3N/4$; however,
as $N$ gets large $N_{\rm expressed}\to N/2$; compare inset Fig. (2). 
In $D^-<\bar D<D^+$ systems typically are stable but decrease their active network-size, 
$N_{\rm expressed}$. In the range $D_{1/2}<\bar D<D^-$ it becomes increasingly likely to sample unstable systems
until for $\bar D<D_{1/2}$ systems almost certainly become unstable. For large $N$ the transition at $\bar D\sim D^-$ 
gets sharp.

As we lower the degradation rate from $\bar D>D^+$ to $\bar D<D_{1/2}$, $N_{\rm expressed}$ decreases from $N$ to $N/2$. 
As a consequence, multiple critical degradation rate values $D_{\rm crit}$ exist in this range, 
where the dynamics of systems switch from one attractor to another one; we may expect a number of switches proportional to $N/2$. 
Between two adjacent $D_{\rm crit}$ the attractors deforms without changing the expressed network.
The reduced (but still considerable) computational cost
allows us to make some quantitative prediction on how the density $\rho_{\rm attractor}$, the number of critical degradation rates (attractor switches) per unit decay rate and molecular species, behaves. 
We specified the ensemble properties $\delta_D=0$, $\bar k=4$, $x^*_i=10+i/N$, $p=1/2$, and 
an average absolute link weight $\bar A=1$ varying only $\bar D$. In Fig. (2) we show estimates of $\rho_{\rm attractor}$ 
for $D_{\rm crit}$ in the ranges $D_{1/2}<D<D^+$ (blue curve) and $D^-<D<D^+$ (red curve). We obtain the estimate by measuring 
the number of times the expressed network size changes by at least one node $i$ in the respective intervals. 
We have done so for $N=10,15,\cdots,40$, and $N=60$, taking averages over $50$ samples. 
We decreased $\bar D$ in increments of $\Delta D=0.01$, starting from $\bar D=2.2$ down to $0.8$. 
We evolved the systems over $T=300$ time units with a time-increment of $dT=0.1$.
Each time we start run the dynamics for another $T$ time units after adapting $\bar D$ we discard the first $200$ time units
and analyze the dynamics between time unit $200$ and $300$. The attractor density $\rho_{\rm attractor}$ 
in the range $D^-<\bar D<D^+$ grows more strongly with a rate of $\kappa_1\sim0.3676$, while the density in 
$D_{1/2}<D<D^+$ only grows with $\kappa_2\sim0.2278$ attractor switches per unit decay rate and molecular species.
Between $D^-$ and $D^+$ the attractor density grows much stronger than between $D_{1/2}$ and $D^-$. 
This is compatible with the observation that for $N\gg 60$ the typical system dynamics will again 
switch between stable attractors in the interval $\bar D\in[D^-,D^+]$, reported in earlier work, 
and then become instable for $\bar D<D^-$.

$D^+$, $D^-$, and $D_{1/2}$ depend on $\bar k$ and not on $N$. At the same time, the number of critical values 
$D_{\rm crit}$ is proportional to $\rho_{\rm attractor}$ and grows monotonically with $N$; compare Fig. (2). 
As a consequence, as cMRN become large, ever smaller fluctuations of $\bar D$ can trigger switches from one attractor to another one. For $\bar k=4$ and realistic MRN sizes of $N=50000$ molecular species and genes, we estimate about $N(\sqrt{\bar k}-\sqrt{\bar k/2})\rho_{\rm attractor}\sim 10500$ attractor switches between $D^+$ and $D^-$.
It follows, using the correspondence principle, that cells that do not control 
their molecular degradation rates frequently suffer from spontaneous re-differentiation 
triggered by degradation rate fluctuations. 

In order to find out whether in a complex dynamical system degradation rates are still informative about the abundance 
$x_i$ of species $i$, we proceeded as follows: We sampled systems specified by $N=25$, $\bar k=4$, $x^*_i=10$, $\delta D=0.5$, $p=0.5$. 
We took $100$ samples for each value $\bar D$ in the range $1<\bar D<2.2$ in increments $0.05$. For every sample we compute the rank order of the degradation rates. The largest decay rate $D_i$ has rank $r=1$, the second largest, rank $r=2$, etc.
For each rank $r$ we compute the fraction of time molecule types $i$ with $r={\rm rank}(D_i)$ becomes expressed and average this fraction over all samples. Surprisingly, there is no significant correlation between the rank order of $D_i$'s and the likelihood of $i$ to be on or off. 
Also the maximal or average abundance of $i$ does not significantly correlate with the rank of $D_i$. 
This unexpected observation again demonstrates that intuitions developed in simple systems can fail us in complex dynamical situations.
The choice of $\bar D$ and $\delta D$ becomes more important for the characteristics of systems dynamics than any individual 
$D_i$ for the abundance of $i$. As a consequence we have to acknowledge that gaining some control over the dynamics of 
large cMRN is a difficult task and inducing unintended results will be a frequent consequence of such attempts.

\subsection{Robustness}

In order to understand the robustness of the systemic behavior discussed above, 
under variations of the network topology, one needs to understand the underlying mathematical principle 
that relates degradation rate changes to switches between attractors, Girko's law, which slightly generalized states the following:\\
For adjacency matrices $A$ with average connectivity $\bar k$, with non zero $A_{ij}$ distributed with mean zero and variance $\sigma_A$, the eigenvalues of $A$ are evenly distributed in a circle of radius $R=\sigma_A\sqrt{\bar k}$. -
Moreover, the matrix, $A'_{ij}=A_{ij}-\bar D \delta_{ij}$, has eigenvalues $\eta'=\eta-\bar D$, where $\eta$ 
are the eigenvalues of $A$.

Since the maximal real eigenvalue $\lambda\sim\sigma_A\sqrt{\bar k}-\bar D$ (Lyapunov exponent) determines whether the dynamics with respect to $A'$ is stable ($\lambda<0$) or not ($\lambda>0$), one can estimate $D^+$ using $0\sim\sigma_A\sqrt{\bar k}-D^+$.
As a consequence, the PC Eq. (\ref{poscond}) becomes relevant once $\bar D<D^+$ and systems start to switch to smaller expressed networks, $N_{\rm expressed}<N$, with $\lambda\leq 0$. The expressed subsystem again has a spectrum of eigenvalues. By varying $\bar D$ further, the subsystem will again become instable and another attractor emerges. This can be iterated until the system gets irrecoverably unstable. 

This self-similar, iterative mechanism does not rely on the exact conditions of Girko's law. 
What is required for the mathematical mechanisms to work qualitatively as described above 
is that the eigenvalues of the weighted adjacency matrix $A$ remain bounded in the complex plane, and 
the value of the maximal real eigenvalue of the systems adjacency matrix $A$ remains sensitive to changes in 
the average degradation rates $\bar D$. 
Fig. (4) shows the eigenvalue spectrum of adjacency matrices with $N=200$, $\bar k=10$. 
The $\bar k N$ non-zero entries in $A$ are set to $A_{ij}=1$ with probability $p$ and to $A_{ij}=-1$ with probability $1-p$, for $p=0.2,0.5,0.8$. Also shown, the spectra for $\bar D=0,4,8$. Further, we consider degradation rates $D_i$ to be evenly distributed in the intervals $\bar D(1-\delta_D)<D_i<\bar D(1+\delta_D)$ for $\delta_D=0,0.5,1$.

For $p>0.5$ a single real maximal eigenvalue emerges at the right side of the circle 
(corresponding to the Perron-Frobenius eigenvalue of positive matrices).
This eigenvalue decreases the value of $D^+$. 
As a consequence, systems with $p>0.5$ will be more likely to become instable 
in general, since also subsystems are likely to inherit $p>0.5$. 
On the other hand, if $p<0.5$ becomes sufficiently small, then this single minimal real eigenvalue emerges to the left of the circle. This eigenvalue can not alter $D^+$. This indicates that cMRN that evolve to remain stable, will experience an evolutionary pressure that favors $p\leq 0.5$ in cMRN.
   
Similarly, if $\delta_D=0$, then the circle is shifted as a whole by $\bar D$. If $\delta_D>0$, increasing $\bar D$ stretches the circle along the real axis. While the smaller real parts of the eigenvalues remain sensitive to $\bar D$, the eigenvalues with larger real parts show a decreasing sensitive to changes in $\bar D$, until for $\delta_D=1$, 
always some $D_i\sim 0$ and the maximal real eigenvalue can no longer become lower than zero. 
Regulatory molecules that do not degrade spontaneously make systems vulnerable to instabilities and it becomes impossible for the cMRN to express the entire network simultaneously. Active degradation mechanisms become necessary to restore dynamical stability.
%
\begin{figure}[th]
	\centering
		\includegraphics[width=8cm]{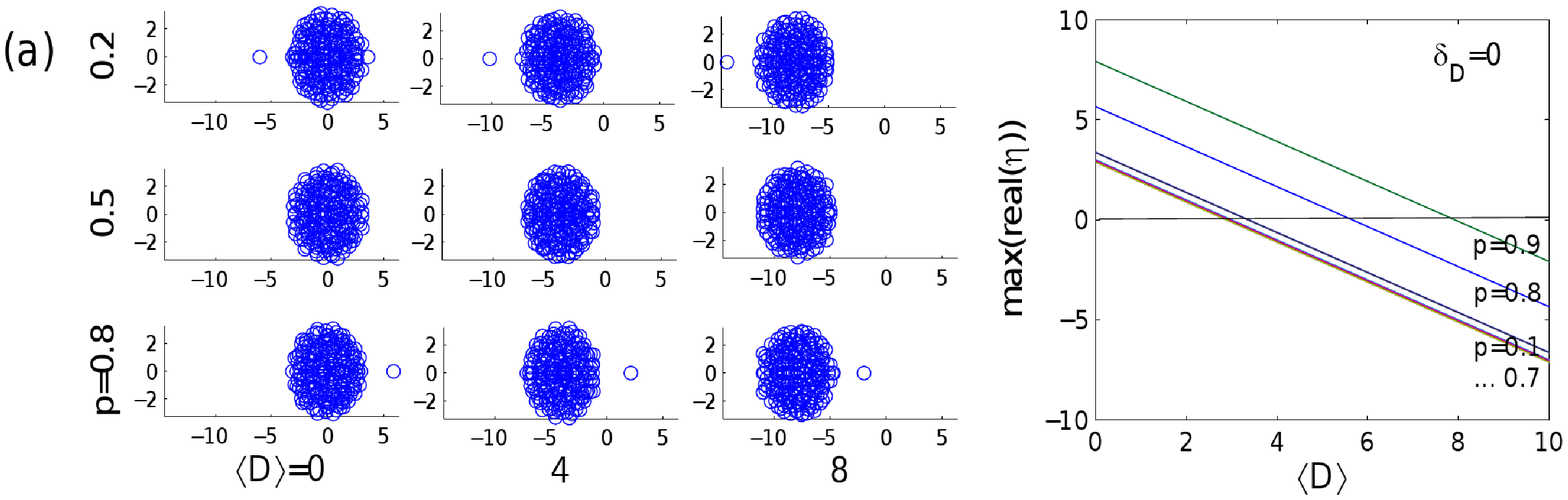}\\
		\includegraphics[width=8cm]{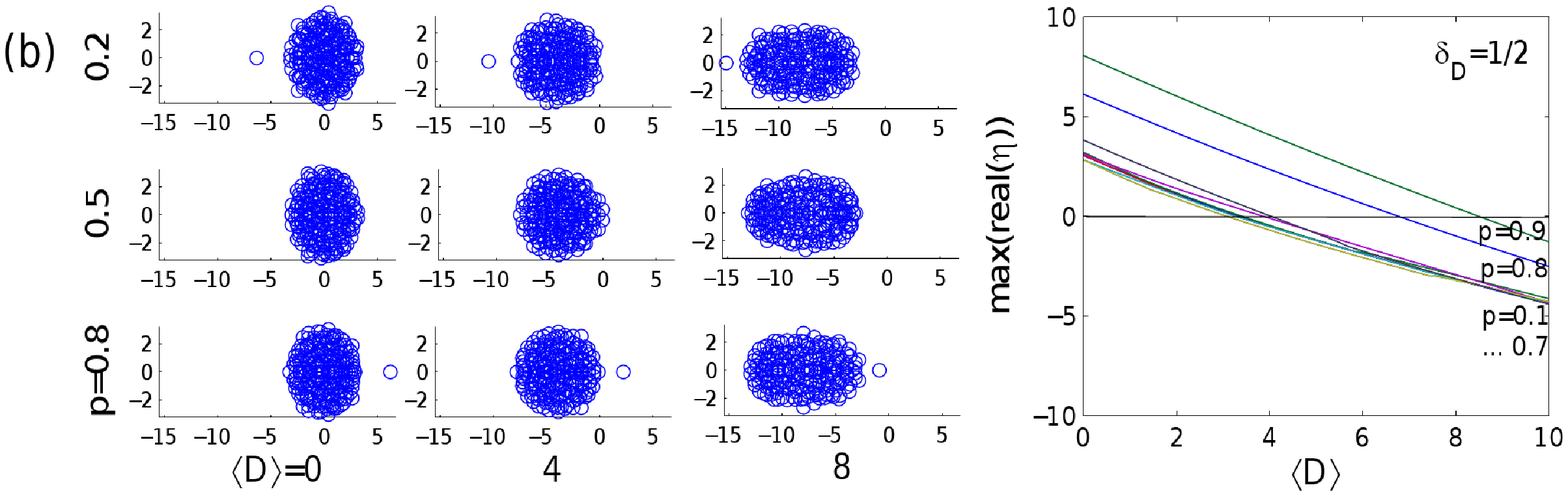}\\
		\includegraphics[width=8cm]{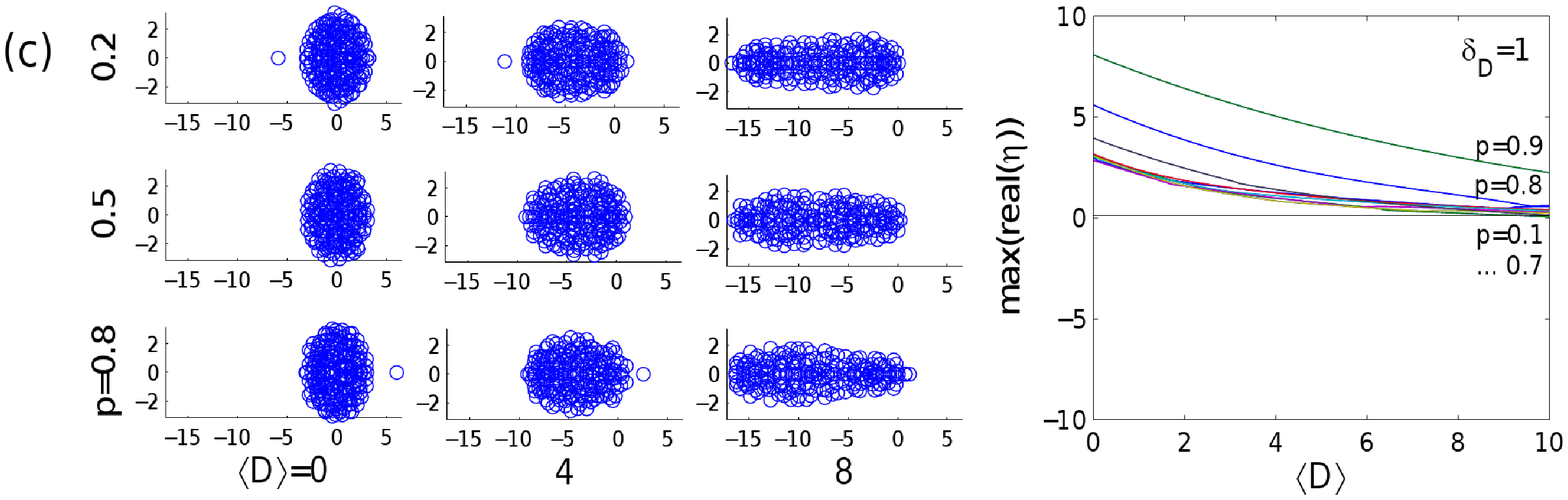}\\
	\caption{{\bf Robustness}
	We demonstrate the effect of the ratio between number of positive entries $A_{ij}=1$ and negative entries $A_{ij}=-1$ 
	in the interaction matrix $A$ for Erd{\"os}-R{\'e}nyi networks. 
	The number of positive entries is given by $p\bar k$ and the number of negative entries is given 
	by $(1-p)\bar k$. The number of agents used in all panes is $N=200$ with an average node degree of $\bar k=10$. 
	The left panes show the eigenvalues $\eta$ of respective matrices $A-D$ in the complex plane, with $\bar D(1-\delta_D)<D_i<\bar D(1+\delta_D)$ for 
	the average degradation rates $\langle D \rangle=\bar D$ and 
	(a) $\delta_D=0$, 
	(b) $\delta_D=0.5$, 
	(c) $\delta_D=1$, 
	for values of $p=0.2,0.5,0.8$ and $\bar D=0,4,8$.
	The panes to the right shows how the maximal real eigenvalue $\lambda=\max(\real(\eta))$ of $A-D$ for $\delta_D=0,0.5,1$ depends on $\bar D$.
	We note that for $\delta_D=1$ and therefore some $D_i\sim 0$ it becomes impossible to shift the maximal real eigenvalue $\lambda$
	below zero. Instead $\lambda\stackrel{>}{\sim} 0$ for all $\bar D$ and the eigenvalue spectrum becomes stretched out as
	one increases $\bar D$. 
	}
	\label{fig:robust}
\end{figure}
%
\section{Conclusions}
Mathematical models of (non-linear) dynamical systems cease to be good predictors of the {\em exact} dynamics if systems become large and complex. This is true for astronomers who want to understand the evolution of galaxies and not the simple Newtonian two body problem, and this is true for systems biology where we want to understand the molecular ``behavior'' of cells, organisms, or eco-systems and not some isolated chemical reaction.
Our ignorance and lack of controll of initial conditions, exact parameter values etc.; 
all the known and unknown unknowns of particular reaction networks, non-linearities of the dynamics and noise, all together render the dynamical systems approach for exact predictive modeling impractical and unreliable. 
However, by considering ensembles of models sharing the same known properties we can still 
obtain useful quantitative information about {\em typical} systemic behavior and stability. 

By identifying MRN with mathematical models of non-linear dynamical systems, different attractors of MRN represent cells with different differentiation status. We show quantitatively that for dominantly catalytic systems the sensitivity of the system dynamics to variations of the average degradation rates increases with system size. Small variations of the average degradation rates can cause cell re-differentiation, i.e. a possibly severe restructuring of the attractors of MRN dynamics. Without an active degradation rate control system, large MRN would frequently re-differentiate their expressed networks spontaneously. In this sense, differentiation comes for free, but not systemic stability. 

The catalytic backbone of reactions in a cMRN emerges if the system is stable enough for the dynamics to approach a specific attractor, which processes most of the available substrate. As a consequence, systemic stability gives catalytic sets within the MRN an advantage in perpetuating themselves (analogous to proliferative fitness) by claiming most of the available resources. 
Therefore, cMRN selection for systemic stability can be expected to experience an evolutionary pressure that favors 
(i) MRNs containing no less suppressor than enhancer links,
(ii) degradation rates distributed narrowly around their mean,
(iii) degradation rates $D_i$ to be all sufficiently bounded away from zero, and 
(iv) the emergence of mechanisms for actively degrading stable functional molecules such as proteins. 

Proteasomes are evolutionary old proteases that provide a 
general mechanism for degrading proteins without exposing the interior of a cell randomly to proteolytic forces. 
The ubiquitous demand on such general mechanism may introduce fluctuations in their general 
availability (e.g. compare \cite{Dantuma2006}). Logistic limitations of proteasome availability 
may cause an overall decrease of degradation rates if a cell under stress increases its overall 
proteolytic demands. Therefore, DRC systems need to appear very early in the process of chemical evolution
to prevent detrimental switching between attractors from happening. 
The simplest way for CDC to emerge then can be understood as a consequence of evolution entangling DRC mechanisms tightly 
with tasks of suppressing unfavorable switches between attractors, i.e. differentiation events. 
Such evolutionary plasticity of degradation rates may partly be explained by the surprising finding that in a systemic context the rank of 
degradation rates ceases to be informative on the abundance of the corresponding molecules.

The presented approach provides us with the exciting possibility to study the effects of MRN topology, such as compartmentalization of the MRN, on system dynamics and stability in a top down fashion, allowing us to identify probable drivers of systemic evolution.
Moreover, the role degradation mechanisms play for the stability of molecular regulatory networks strikingly reminds us of questions of sustainability in other complex regulatory systems. For instance, in ecological or economical systems there need to exist features analogous to DRC and CDC systems in cells, necessary for sustainable {\em and} stable system dynamics.
\vfill 




\end{document}